\documentclass{emulateapj}

\newcommand{\mybold}{}

\slugcomment{To appear in ApJ.}

\shorttitle{Star Formation in PKS 2005-489}
\shortauthors{Bressan et al.}

\begin{document}

\title{Ongoing Star Formation in the BL Lacertae object PKS 
2005-489}
\author{A.Bressan$^{1,2,3}$, R.Falomo$^{1}$, 
J.R. Vald\'es$^{3}$, R.Rampazzo$^{1}$}

\affil{$^{1}$ INAF Osservatorio Astronomico di Padova, vicolo dell'Osservatorio 5, 35122 Padova, Italy\\
$^{2}$ SISSA, via Beirut 4, 34014, Trieste, Italy\\
$^{3}$ Instituto Nacional de Astrof\'{\i}sica, Optica y Electr\'onica, Apdos. Postales 51 y 216, C.P. 72000 Puebla, Pue., M\'exico}

\email{bressan@pd.astro.it}
\email{falomo@pd.astro.it}
\email{jvaldes@inaoep.mx}
\email{rampazzo@pd.astro.it}

\begin{abstract}
We present  VLT long slit optical spectroscopy of the 
luminous BL Lacertae object PKS 2005-489.
The high signal-to-noise ratio and the good
spatial resolution of the data allow us to detect the signatures
of ongoing star formation in an extended rotating ring, at $\sim$ 4 kpc
from the nucleus. We find that the ring is almost perpendicular to the
radio axis and its total star formation rate is $\simeq$ 1 M$_\odot$/yr. 
We briefly discuss the concomitant presence of recent star
formation and nuclear activity.
\end{abstract}


\section{Introduction}
The demographics study of super massive black holes (SMBH) in galaxy centres
(e.g. Combes 2005 and references therein) 
and the similarity between the strong cosmic evolution of
active galactic nuclei (AGNs) and the cosmic star formation rate (SFR)
\citep{steidel99,fan01},
lead to the coevolution concept, where 
AGN evolution traces, if not regulates,
the build up of the spheroids \citep{granato04}.
In this scenario the growth of the central massive black hole (SMBH) 
and the star formation rate are mutually dependent through the 
energetic feedback until, in the most massive objects, the
SMBH shines at power sufficient to evaporate all the 
gaseous component. Thereafter the galaxy evolves passively,
eventually undergoing sporadic episodes of SF episodes 
that likely accompany the
acquisition of fresh gas during interactions.
Effects of such recent rejuvenation episodes are often seen as 
prominent H absorption features (typical of A stars)
in the optical spectra of early type galaxies \citep{Longhetti1999,
Aretxaga2001}, 
or more recently in their mid infrared spectral region 
\citep{Bressan06, pan06}.
A number of observational evidences for a tight link between
star formation and nuclear activity have been reported 
\citep{Bressan02, heckman04, DellaValle05, canalizo06}.
However, whether these rejuvenation episodes are triggered
by a reactivation of the SMBH (Ho 2005), 
or the reactivation is only concomitant with the SFR,
still remain a very debated issue. 

To cast light on this point, many imaging studies have been performed in
recent years on the host galaxies of nearby and distant AGNs 
\citep{bahcall97,dunlop03,pagani03,falomo04,falomo05,kukula01}.
For nearby sources (z $<$ 0.5) these studies have
depicted a picture that indicates that radio loud quasars (RLQ) are
preferentially found in luminous ellipticals exceeding by $\sim$1--2 mag the
typical galaxy luminosity L$^*$ (M$^*_H \sim$ --23.8; Mobasher et al. 
1993)\footnote{We adopt H$_0$=70, $\Omega_L$=0.7 and $\Omega_M$=0.3}, 
while radio quiet quasars (RQQ) can be
hosted both by ellipticals and spirals of somewhat lower luminosity (e.g.
Percival et al. 2001) but with a clear tendency to be
ellipticals for high luminous RQQs.
For BL Lac objects (BLL), characterized by nuclear luminosities $\sim$ 5-10
smaller than that of RLQs, it is found that the host galaxies
are unperturbed giant ellipticals, 
with luminosity comparable  to that of the RLQs hosts 
(Urry et al 2000; Falomo, Carangelo \& Treves, 2003a).

All imaging studies are, however, unable to address the issue of stellar
content of these galaxies (apart from a very preliminary insight through
multicolor images (e.g. Jahnke et al 2001; Labiano et al 2005a).  The only
effective way to investigate the stellar population of the host galaxies is
using adequate spectroscopy of the surrounding nebulosity.
Till now only pioneering work has been done on
this front (e.g. Boroson et al 1985; Nolan et al 2001) 
or somewhat more detailed study on
individual sources (e.g. Canalizo \& Stockton 2000, Labiano et al 2005b).

In order to investigate the link between nuclear activity and star formation
we are carrying out  a {\mybold spatially resolved} 
spectroscopic study of the host galaxies of low and
high luminosity nearby AGN. In this letter we report on the discovery of a
rotating star forming ring in PKS 2005-489, one of the brightest BL Lac
objects of the southern emisphere.

\section{Observations and Data Analysis}
\label{observations}
The radio source PKS 2005-489 was detected in the Perkes 2.7 GHz
survey \citep{wall75}, subsequently identified by \citet{
savage77} as an N galaxy and then classified as a BL Lacertae
object by \citet{wall86}. The redshift
of the object (z=0.071) was derived by \citet{falomo87} on the
basis of two faint emission lines identified as H$\alpha$
and NII 6584. 
The luminous nucleus (m$_{R}$=12.7) is hosted by a massive and
luminous spheroidal galaxy of m$_{R}$=14.5 and half light radius
r$_e$=5.7\arcsec\ \citep{urry2000}.
This galaxy (M$_{R}$=-23.1; R$_e\sim$9 kpc) is the dominant member of 
a poor group of galaxies
with the closest object at 65 kpc (Pesce et al. 1994).

We obtained high signal to noise (S/N  $\sim$ 250) 
optical spectra of PKS 2005-489
using the Focal Reducer and low dispersion Spectrograph (
FORS1, \citet{app}) at the Very Large Telescope (VLT) of the European
Southern Observatory (ESO).
Two spectra of 1200 seconds were secured under good seeing conditions 
($\sim$1\arcsec) in August 2002, 
using the grism GRIS600V (dispersion 49 A/mm) and
a 1\arcsec\ wide long slit
centered in the nucleus and oriented in the North-South direction.
The spectra cover the 4840 and 7200 \AA\ wavelength range with 4.8\AA\
spectral resolution.

The plate scale on the detector is 0.20 
arcsec/pixel\footnote{At the redshift of PKS~2005-489,
the distance is 321 Mpc and 0.2 arcsecs correspond to 0.31 kpc.}.
Standard data reduction (including flat fielding, cosmic ray rejection
and calibration) was performed using IRAF tasks. Spectra were
wavelength-calibrated using  He-Ar lamp observations with an accuracy of
0.2\AA. Finally absolute flux calibration was provided by observations
of the standard star LTT6248.
The combined spectrum reaches a S/N of $\sim$250 in the nuclear region
and S/N of $\sim$50 at 3\arcsec\ from the centre.

%
\begin{figure} \begin{center}
\includegraphics[width=0.45\textwidth,height=0.4\textheight]{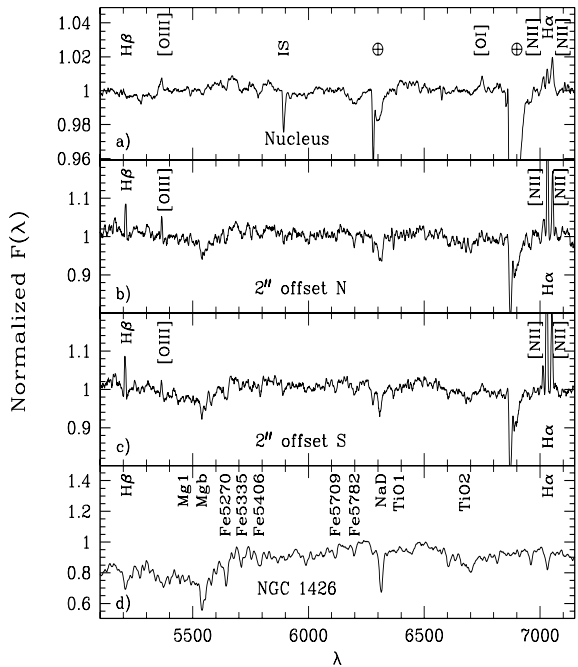} 
\caption{Spectra of PKS 2005-489. Panel a) is for the nuclear region while panels b)
and c) {\mybold refer to two symmetric off-nuclear 
regions at $\sim$ 2\arcsec\ from the center.}
For comparison the spectrum of the early type galaxy NGC 1426 
(Rampazzo et al 2005) is shown in panel d). Main
emission and absorption features are labelled.
\label{fig1}    } 
\end{center} 
\end{figure}
\label{results}
{\mybold In Figure~\ref{fig1} we show the comparison of the spectrum of PKS 2005-489
extracted in the nuclear region with that obtained from two
regions at $\sim$ 2 arcesc from the nucleus.
While the nuclear spectrum  is dominated by the non thermal emission, 
a significant contribution from the host
galaxy is clearly visible  in the off nuclear regions.}

In all these spectra a number of
narrow emission lines, due to HI, O[I], O[III] and N[II] 
are also clearly visible. 
{\mybold Their detection in the extra nuclear spectra
suggests} the presence of an extended 
line emission region around the nucleus.

\section{ The nature of the extended line emission region}
\begin{figure}
\centering
\includegraphics[width=0.3\textwidth, angle=90]{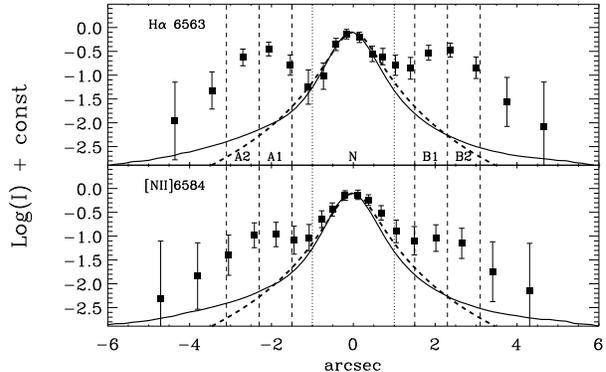}
\caption{Spatial intensity profiles (filled squares)
of H$\alpha$ (upper panel) and [\ion{N}{2}]$\lambda$6583\AA\ 
(lower panel) in PKS 2005--489.
The solid and dashed lines represent the spatial profile in the adjacent continuum
and the PSF derived from a stellar spectrum, respectively. All the profiles are normalized
to the peak of the continuum profile. At the assumed distance of the object (320 Mpc), 
1\arcsec\ corresponds to 1.5 kpc. {\mybold The vertical dotted lines delimit the nuclear region (N)
while the vertical dashed lines delimit the off-nuclear regions (see Table 1).}
\label{fig2}}
\end{figure}

In order to characterize the properties the extended emission line region
we compared the spatial distribution 
of the line intensity ($H\alpha$ and N[II]6584) with
that of the adjacent continuum (Figure~\ref{fig2}).
For both emission line we extracted the spatial intensity profile in
bins of variable width, from 0.2\arcsec\, in the centre, to 1\arcsec\ at 
5\arcsec\ from the nucleus. 
The shape of the spatial profile of the emission line
intensity is consistent with that of the continuum emission and of the 
point spread function (PSF)
out to a radius of $\sim$ 1.5\arcsec. Beyond this radius it exhibits a marked
excess that peaks at about 2.5\arcsec\ (corresponding to $\sim$ 4 kpc). 
Both H$\alpha$ and 
[\ion{N}{2}]$\lambda$ 6583 \AA\ show spatial profiles with similar shape.
The comparison clearly indicates a resolved 
extra nuclear emission region. 

This region  is sufficiently close to the nucleus that photoionization of gas
from the nuclear emission could be responsible for the observed emission.
Alternatively the emission could arise
in the gas associated to a star forming region and ionized by hot stars.
%

\begin{figure}
\begin{center}
\includegraphics[width=0.4\textwidth]{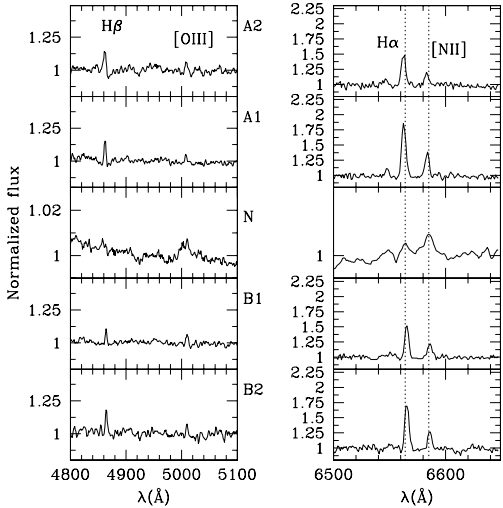}
\caption{
Normalized rest-frame optical spectra of the nuclear and circum-nuclear
regions of the BL Lac object PKS 2005-489. Left panels: The H$\beta$ [OII]
5007 spectral range. Right panel: The H$\alpha$ and [\ion{N}{2}]$\lambda$6584
range.
\label{sppks}}
\end{center}
\end{figure}

\begin{figure}
\centering{
\includegraphics[width=0.32\textwidth]{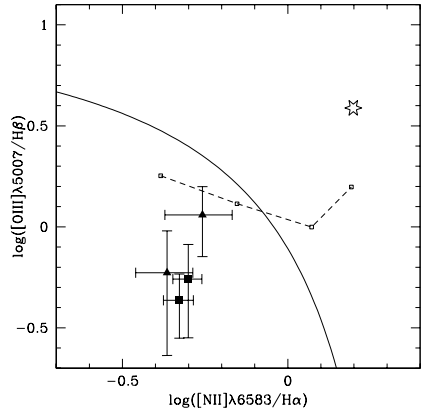}
}
\caption{
Diagnostic diagram (of [\ion{O}{3}]$\lambda$5007/H$\beta$ vs [\ion{N}{2}]$\lambda$6583/H$\alpha$) for
the nuclear (star) and  the four circum-nuclear regions of the BL Lacertae object PKS 2005-489. Filled
squares refer to zone A while filled triangles refer to zone B. 
The transition between the characteristic
starburst emission region and AGN-Liners emission region is indicated by the solid line 
(Kewley et al. 2001), while the dashed line connects "shock only" models with
B/n$^{1/2}$=2 $\mu$G cm$^{3/2}$ computed
by \citep{dopsut96}.
Emission line ratios
are corrected for Galactic and internal extinction.
\label{diadia}}
\end{figure}

The nature of the emission can be disclosed by means of  the classical
diagnostic diagram based on
the [\ion{O}{3}]$\lambda$ 5007/H$\beta$ and [\ion{N}{2}]$\lambda$6583/H
$\alpha$ line intensity ratios \citep{Baldwin81}.
We extracted the average spectrum of the
nucleus (aperture of 2$\arcsec$ diameter) and in four off-nuclear regions, 0.8$\arcsec$ wide,
centered at 2.7$\arcsec$ (A2 and B2)  
and 1.9$\arcsec$ (A1 and B1), South and North from the centre, respectively.
These spectra are shown in  Figure~\ref{sppks} and  the corresponding
emission line intensities are reported in Table \ref{emission}.
The [\ion{N}{2}]$\lambda$6583/H
$\alpha$ vs [\ion{O}{3}]$\lambda$5007/H$\beta$ 
diagnostic diagram for the different regions is shown 
in Figure~\ref{diadia}. We applied a correction for 
galactic extinction of E(B-V)=0.056 \citep{schlegel98}.
The internal
extinction was calculated from the
Balmer decrement, assuming an intrinsic I(H $\alpha$)/I(H$\beta$)~=~3.0,
and adopting the extinction law of \citet{cal00}. The value of the internal
extinction in the nuclear region is E$_{(B-V)}\simeq$1.3, while in the
circum-nuclear regions is E$_{(B-V)}\simeq$0.4.

The nucleus clearly occupies the region of gas photoionized by non thermal
emission. By contrast the line intensity ratios
of the  extended emission regions, are 
typical of HII regions or starbursts \citep{veilleux87, Kewleyetal2001}.
A significant contribution by shocks seems also ruled out
(see also Figure~2b of Dopita \& Sutherland, 1996).
Accounting for the contribution of old stars 
to the H$\alpha$ and H$\beta$ intensities in the extended regions,
would render our conclusion even more robust.

%
\begin{deluxetable}{lrrrrrr}
\tablecaption{Observed Emission line Intensities of PKS 2005-489
\label{emission}}
\tablewidth{0pt}
\tabletypesize{\scriptsize}
\tablehead{\colhead{Zone} 
& \multicolumn{1}{c}{r} 
& \multicolumn{1}{c}{$\Delta$r} 
& \multicolumn{1}{c}{H$\beta$} 
& \multicolumn{1}{c}{[\ion{O}{3}]} 
& \multicolumn{1}{c}{H$\alpha$} 
& \multicolumn{1}{c}{[\ion{N}{2}]}\\
\colhead{} 
& \multicolumn{1}{c}{(\arcsec)} 
& \multicolumn{1}{c}{(\arcsec)} 
& \multicolumn{1}{c}{$\lambda$4861} 
& \multicolumn{1}{c}{$\lambda$5007} 
& \multicolumn{1}{c}{$\lambda$6563} 
& \multicolumn{1}{c}{$\lambda$6583}
}
\startdata
A2 & 2.7& 0.8 & 2.5$\pm$0.8 & 1.5$\pm$0.7  & 10.6$\pm$1.1 & 4.6$\pm$0.8\\
A1 & 1.9& 0.8 & 2.3$\pm$0.3 & 2.8$\pm$0.9  & 11.7$\pm$1.4 & 6.5$\pm$1.1\\ 
N  & 0  & 2.0 & 16$\pm$11   & 71$\pm$14    & 204$\pm$35   & 327$\pm$35\\
B1 & 1.9& 0.8 & 4.0$\pm$1.2 & 2.3$\pm$0.8  & 19.0$\pm$1.8 & 9.5$\pm$0.7\\
B2 & 2.7& 0.8 & 2.8$\pm$0.6 & 1.3$\pm$0.3  & 13.6$\pm$0.8 & 6.4$\pm$0.5\\
\enddata
\tablenotetext{a}{Intensities are in units of 10$^{-17}$ erg s$^{-1}$ cm$^{-2}$}
\end{deluxetable}
\begin{figure}
\centering
\includegraphics[width=0.32\textwidth,angle=270]{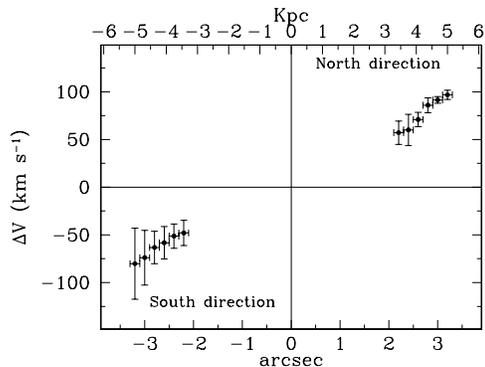}
\caption{Velocity profile, in Km/s, obtained from the mean
displacement of the H$\alpha$ and [\ion{N}{2}]$\lambda$6583 \AA\ line centres
along the slit.
The lower horizontal axis is in arcsecs while 
the upper horizontal axis is in kpc.
\label{rotcurve}
}
\end{figure}
An additional relevant information to understand the nature
of the extended line emission region is provided by the kinematics.
In fact, simple inspection of Figure~\ref{sppks}
shows a progressive shift of the centre of
H$\alpha$ and [\ion{N}{2}]$\lambda$6583\AA\ along the slit direction, 
by $\Delta\lambda\simeq$2\AA. 
We checked the reality of this displacement by comparison with that
of atmospheric features, close to the emission lines, 
that exhibit a maximum shift of 
$\Delta\lambda\simeq$0.06\AA.

The mean shift of the H$\alpha$ and [\ion{N}{2}]$\lambda$6583\AA\ lines,
at various positions from the centre, is reported in Figure~\ref{rotcurve}. 
The shape of the wavelength shift together with the
symmetry of the spatial profile of the emission (see Figure~\ref{fig2}),
suggest the presence of a rotating ring around the nucleus 
located at about  4 kpc from the centre. 

\section{Discussion}
\label{conc}
We have reported on high signal to noise spectroscopy 
of the BL Lac object PKS 2005-489 that reveals
the presence of an extended emission region around the nucleus. The
spatial intensity profile of the line emission regions, together with the
kinematics of the gas, indicate the presence of a rotating ring with ongoing
star formation at $\sim$4 kpc from the nucleus.

We estimate that the star formation rate in the ring is SFR
$\sim$ 1.2 M$_{\odot}$ yr$^{-1}$. This is based on the
relationship (Panuzzo et al. 2003)
 between the SFR and 
the extinction corrected H$\alpha$ luminosity, assuming that 
the signal in our spectra (integrated in two slices of 1$\arcsec\times$1.6$\arcsec$)
samples $\sim$1/7 of the entire ring.

Assuming a constant SFR in the last 100 Myr, a
typical lifetime for small starbursts \citep{pan06},
the mass associated with this  recent burst is
M$_{burst}\simeq$10$^8$M$_{\odot}$. 
This rejuvenation episode involves only 
the 0.03\% of the host galaxy mass (M$_{host}\simeq$4.2~10$^{11}$M$_{\odot}$,
estimated from the absolute R luminosity).
Since this is about  hundred times smaller than the typical fractional mass
($\sim$ a few percent) 
involved in episodes of recent star formation 
(e.g. Longhetti et al. 1999, Annibali et al. 2006), it is unlikely
that such rejuvenation episodes are 
picked up  in quiescent ellipticals of similar luminosity.

The value of the  SFR is comparable to the accretion rate onto
the SMBH, $\dot{M}\simeq$0.8 M$_{\odot}$ yr$^{-1}$,
estimated  assuming a $\dot{M}/\dot{M}_{Edd}\simeq$0.047 (Xie et al. 2004),
a black hole mass of Log(M$_{BH}$)=8.9 \citep{falomo03a} and an 
efficiency $\eta$=0.1.

Using the relation between the circular velocity and
the central velocity dispersion \citep{pizzella05}
the intrinsic velocity of the ring should be
$\sim$ 420 km/s (adopting $\sigma_c$=250 km/s for a galaxy with this luminosity).
Since the observed  averaged radial velocity of the ring is V$_r\sim$75 km/s
the inclination angle of the ring is $\sim$80 degrees.
Therefore the star forming ring is seen almost face on.
{\mybold Since, according to the unification scheme of radio loud
AGNs (Urry \& Padovani, 1995)
BL Lac objects have the radio jet aligned along the line
of sight to the observer, the star forming ring
is nearly perpendicular to the axis of the jet.
A similar case was reported for the powerful radio galaxy 3C218, (Hydra A)
where a fast rotating ($\sim$450 km/s) disk of young stars and gas 
of few kpc was found (Melnick et al. 1997). Also in this case the disk 
is found perpendicular to the position of the radio jet.}

These observations suggest a link between  disk like star forming regions
and the nuclear activity.
The presence of the rotating star forming region
is the signature of a minor acquisition event
that has fuelled of gas the central few kpc region of PKS 2005-489. 
The gas had time to reach a rotating stable configuration and to begin
the star formation process (e.g. Wada 2004 and references therein).   
Since the ring is almost orthogonal to the radio axis it is unlikely
that the star formation has been induced by the jet.
Moreover the observed geometry also
suggests that the nucleus has been fuelled and activated by
gas associated with the same event, 
after a further significant loss of angular momentum.

\acknowledgments
We thank the anonymous referee for useful suggestions.
J.R.Valdes is grateful to
OAPD for its warm hospitality.
A. Bressan acknowledges worm hospitality by INAOE.
This work is based on observations collected at the 
European Southern Observatory, Paranal, Chile.

\end{document}